\documentclass[twocolumn,prl,showpacs]{revtex4}

\usepackage{graphicx}

\begin{document}

\title{Fundamental limitations to gain enhancement in periodic media and waveguides}%

\author{Jure Grgi\'{c},$^{1}$ Johan Raunkj{\ae}r Ott,$^{1}$ Fengwen Wang,$^{2}$ Ole Sigmund,$^{2}$ Antti-Pekka Jauho,$^{3}$ Jesper M{\o}rk,$^{1}$  and N. Asger  Mortensen$^{1}$\footnote{asger@mailaps.org}}
\address{$^{1}$DTU Fotonik, Department of Photonics Engineering, Technical University of Denmark, DK-2800 Kongens Lyngby, Denmark\\
$^{2}$DTU Mekanik, Department of Mechanical Engineering, Solid Mechanics, Technical University of Denmark, DK-2800 Kongens Lyngby, Denmark\\
$^{3}$DTU Nanotech, Department of Micro- and Nanotechnology, Technical University of Denmark, DK-2800 Kongens Lyngby, Denmark}

\date{\today}

\begin{abstract}
A common strategy to compensate for losses in optical nanostructures is to add gain material in the system. By exploiting slow-light effects it is expected that the gain may be enhanced beyond its bulk value. Here we show that this route cannot be followed uncritically: inclusion of gain inevitably modifies the underlying dispersion law, and thereby may degrade the slow-light properties underlying the device operation and the anticipated gain enhancement itself. This degradation is generic; we demonstrate it for three different systems of current interest (coupled resonator optical waveguides, Bragg stacks, and photonic crystal waveguides).  Nevertheless, a small amount of added gain may be beneficial.
\end{abstract}


\maketitle

Light-matter interactions in periodic structures can be significantly enhanced in the presence of slow-light propagation. This paradigm has led to several important discoveries and demonstrations, including the enhancement of nonlinear effects~\cite{Soljacic:02,Soljacic:2004,MonatRewPer,Corcoran:2009,Colman:2010,Shinkawa:2011,Boyd:2011}, Purcell effects for light emission~\cite{Lecamp:2007}, light localization~\cite{Sapienza:2010}, as well as slow-light enhanced absorption and gain processes~\cite{Baba:2008, Mortensen:2007,Mork:10,Sakoda1999,Krauss:2007}. Loss is an inherent part of any passive optical material, and the inclusion of gain material is presently receiving widespread attention in many different situations, ranging from the fundamental interest in gain-compensation of inherently lossy metamaterials~\cite{Xiao:2010,Wuestner:2010,Hamm:2011,Stockman:2011} and spasing in plasmonic nanostructures \cite{Oulton:2009,Noginov:2009}, to active nanophotonic devices such as low-threshold lasers~\cite{Masuo:2010} and miniaturized optical amplifiers. There is a common expectation that if a material with net gain $g_0$ is incorporated in a periodic medium, such as Bragg stacks, photonic crystals (PhC) or metamaterials, the gain will effectively be enhanced to $g_{\rm{eff}}\sim n_g^0 g_0$, where $n_g^0$ is the group index associated with the underlying dispersion relation $\omega_0(k)$ of the passive structure. In a device context the gain enhancement is anticipated to allow shrinking the structure by a factor equivalent to the group index, while maintaining the same output performance. However, this reasoning implicitly assumes that gain can be added  without considering its impact on $\omega_0(k)$ -- an assumption that calls for a closer scrutiny.

In this Letter we analyze the modification of the dispersion due to gain, and show that a large gain will eventually jeopardize the desired slow-light dispersion supported by the periodic system, thus suppressing the slow-light induced light-matter interaction enhancement anticipated in the first place. On the other hand, a small amount of material gain is shown to beneficial.  Thus, importantly, devices employing quantum-dot gain material may display a superior performance.

\begin{figure}[b!]
\includegraphics[width=0.45\textwidth]{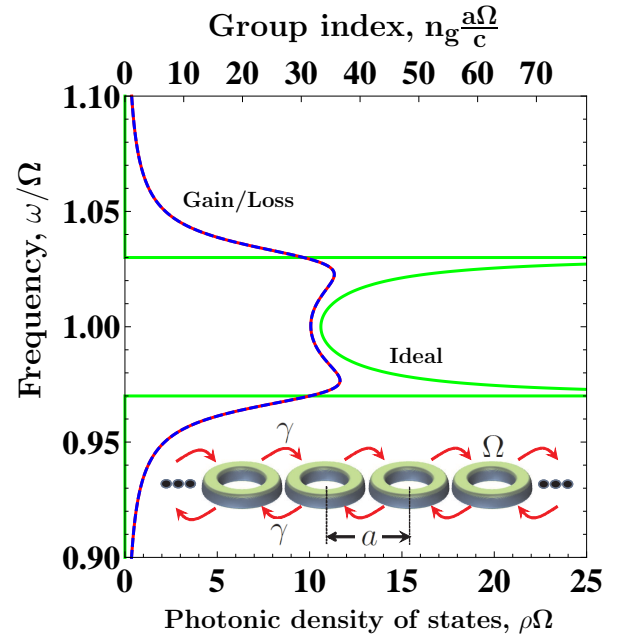}
\caption{(Color online) Photonic density of states (per resonator) $\rho$ (lower horizontal axis) and group index $n_g$ (upper horizontal axis) versus frequency $\omega$, for a CROW with $\gamma=0.03$. For passive resonators with $g_0=0$, van~Hove singularities appear at the band edges (green line). For $g_0=\pm 0.01$, gain (blue-dashed line) or an equivalent loss (red line) cause a similar smearing of the singularities.}
\label{fig:PDOS}
\end{figure}

Early investigations emphasized simple one-dimensional periodic media such as Bragg stacks in the context of slow-light enhanced gain and low-threshold band-edge lasing~\cite{Dowling:1994}. Likewise, the related phenomenon of slow-light enhanced absorption was proposed as a route to miniaturized Beer--Lambert sensing devices~\cite{Mortensen:2007}.
Slow-light enhancement thus appears to be a conceptual solution to a wide range of fundamental problems involving inherently weak light-matter interactions or technological challenges calling for miniaturization or enhanced performance. However, recent studies of linear absorption~\cite{Pedersen:2008,Grgic:2010a} suggest that $n_g$ itself is also affected by the presence of loss. Likewise, the gain may also influence $n_g$~\cite{Sukhorukov:2011} and analytical studies of coupled-resonator optical waveguides (CROW) show explicitly that the group index and attenuation have to be treated on an equal footing and in a selfconsistent manner~\cite{Grgic:2011}. Here, we show that the same considerations apply to gain, and illustrate the general consequences with the aid of three examples. Recent studies on random scattering showed that fabrication disorder leads to a loss that increases with the group index \cite{Patterson:2009,OFaolain:2010}. This effect imposes another limitation to the degree of light slow-down that may be useful for the applications. However, in contrast, the effect investigated here is intrinsic, and will impede the performance even of a perfectly regular structure.

\emph{Coupled resonator optical waveguide.} We consider first a CROW formed by a linear chain of identical and weakly coupled neighboring optical resonators (inset of Fig.~1). In the frequency range of interest the individual resonators support a single resonance at $\Omega$ and when coupled together they form a propagating mode with dispersion relation~\cite{Morichetti:2011}
\begin{equation}\label{crow}
\omega(k)= \Omega\left(1-ig_0\right)[1-\gamma\cos(ka)].
\end{equation}
Here, $a$ is the lattice constant while $g_0$ and $\gamma$ are dimensionless parameters representing the material gain and the coupling, respectively. Our sign convention for the gain term is associated with an $\exp{(i\omega t)}$ time-dependence, corresponding to a real-valued frequency relevant for the excitation by a  CW laser source. Inverting Eq.~(\ref{crow}) leads to a complex-valued Bloch vector $k(\omega)=k'(\omega)+ik''(\omega)$. The group velocity is computed from  $v_g=(\partial k'/\partial \omega)^{-1}$. The photonic density of states (PDOS) is in general proportional to the inverse group velocity and in this particular example $\rho=a/(\pi v_g)$.
In Fig.~\ref{fig:PDOS} we show the PDOS for a typical CROW, eg. for a structure working at around the telecom wavelength, $\Omega\sim10^{15}~\rm{s^{-1}}$, the figure corresponds to a lattice constant of $a\sim 300$ nm. For the passive structure with $g_0=0$ (green line) the characteristic van~Hove singularities at the lower and upper band edges are found. In the presence of damping ($g_0<0$) one expects a smearing of the PDOS and broadening of the singularities (red line)~\cite{Grgic:2011}. Intuitively, one might expect that loss compensation by addition of gain material will sharpen the PDOS features, but \emph{a~priori} it is not clear what net-gain ($g_0>0$) will result in. However, with the dispersion relation (\ref{crow}) one can show that changing the sign of $g_0$ causes {\emph no} changes in the PDOS, as is also evident from the plotted results (blue-dashed line). In the context of the intrinsic quality factor $Q_0$ of the resonators we note that $Q_0=1/(2|g_0|)$~\cite{Grgic:2011}, which in the present case corresponds to a $Q_0=500$. Since $n_g\propto \rho$ we conclude that both loss and gain will reduce the maximal achievable group index, in particular near the band edges where the group index would otherwise diverge. For the lossy case this is easily understood in terms of multiple scattering, where even a small imaginary absorption coefficient will eventually cause a dephasing of the otherwise constructive interference leading to a standing-wave formation at the band edges. For gain the situation is very much the same; in this situation the multiply scattered wave components increase in amplitude and eventually prevent the perfect formation of a standing-wave solution. Mathematically, changing the sign of $g_0$ simply corresponds to a complex conjugation of $k(\omega)$, thus rendering the real part and the derived PDOS and group index invariant. This observation clearly illustrates a potential conflict for the anticipated slow-light enhancement of gain if a too high material gain is added. This effect is not special to the CROW as the following two examples demonstrate.

\begin{figure}[t!]
\includegraphics[width=0.45\textwidth]{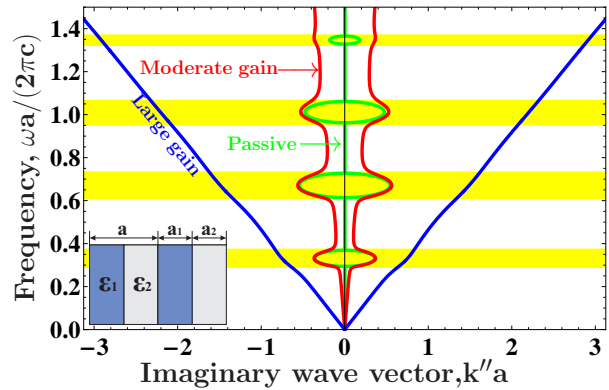}
\caption{(Color online) Imaginary part of Bloch vector $k''$ versus frequency $\omega$, for a Bragg stack with $a_2=2a_1$, $\epsilon_2'=3$, and $\epsilon_1'=1$~\cite{SM}. The passive structure (green line) exhibits clear band-gaps (yellow shading), which are being smeared out for moderate gain/loss, $\epsilon''=\pm 0.1$ (red line). Exaggerated large gain/loss ($\epsilon''=\pm 1$) eventually removes the  band-structure effects (blue line).}
\label{fig:complex}
\end{figure}

\emph{Bragg stack.} Next, we turn to a one-dimensional realization of a more complex PhC concept: the dielectric Bragg stack consisting of alternating layers of thickness $a_1$ and $a_2$, with dielectric constants $\epsilon_1$ and $\epsilon_2$, respectively (inset of Fig. 2). The  dispersion relation is given by
\newpage
\begin{eqnarray}\label{BG}
\cos(ka)&=&\cos\left(\sqrt{\epsilon_1}a_1\frac{\omega}{c}\right)\cos\left(\sqrt{\epsilon_2}a_2\frac{\omega}{c}\right)\\
&&-
\frac{\epsilon_1+\epsilon_2}{2\sqrt{\epsilon_1}\sqrt{\epsilon_2}}\sin\left(\sqrt{\epsilon_1}a_1\frac{\omega}{c}\right)\sin\left(\sqrt{\epsilon_2}a_2\frac{\omega}{c}\right)\nonumber
\end{eqnarray}
where $a=a_1+a_2$ is the lattice constant and $c$ is the speed of light in vacuum. The dielectric constants can  be complex-valued, allowing for analysis of both lossy and gain media~\cite{Shank:1971,Dowling:1994}. The characteristic dispersion diagrams for Bragg stacks are readily derived from $k'(\omega)$. Here we examine the imaginary part $k''(\omega)$, central to our discussion of slow-light gain and loss enhancement. For simplicity, we assume that gain is added to both layers 1 and 2, so that all modes experience the same field overlap with the gain material. Relaxing this assumption will influence the different bands in a slightly different manner, but without changing the overall conclusions. Fig.~\ref{fig:complex} shows a plot of $k''$ versus $\omega$, emphasizing both the positive and negative branches associated with backward and forward propagating branches in the usual $k'$ versus $\omega$ dispersion diagram (not shown, however see Ref.~\onlinecite{SM}). For the gainless material (green line) the imaginary part $k''$ is nonzero only inside the band gaps (shaded areas) while it vanishes inside the bands of free propagation. As the gain is moderately increased (red line) ($g_0\sim2000~$cm$^{-1}$ realizable eg. with GaAs, see \cite{SM}), a finite, enhanced gain develops inside the bands. Clearly,  $k''$ remains finite near the band edges, in contrast to a diverging enhancement as predicted by a lowest-order perturbative treatment~\cite{Mortensen:2007}, where the back-action of material gain on the group index is neglected. For exaggerated larger values of  $g_0$  (blue line) there is no reminiscence of the band gaps:  the structure effectively responds as a homogeneous material.

\begin{figure}[t!]
\includegraphics[width=0.45\textwidth]{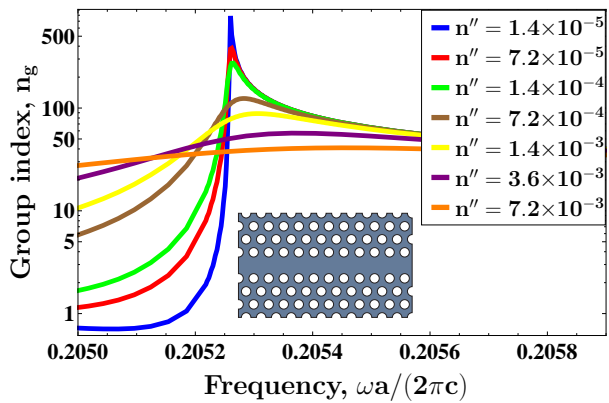}
\caption{(Color online) Group index $n_g$ versus frequency $\omega$, for a photonic crystal semiconductor waveguide with varying gain $g_0\propto n''$.}
\label{fig:ng}
\end{figure}

\emph{Photonic crystal waveguide.} As the final  example, we consider  PhC waveguide structures with a strong transverse guiding due to the presence of a periodic photonic crystal cladding (inset of Fig. 3). Firm light confinement and strong structural dispersion with high $n_g$~\cite{Vlasov:2005,Notomi:2001,Krauss:2007} make such waveguides interesting candidates for compact photonic devices and for fundamental explorations of light-matter interactions~\cite{Sapienza:2010,Mork:10}. Due to the need of a non-perturbative treatment, analytical progress is difficult and we proceed numerically with the aid of a finite-element method. We use a super cell approach with boundary conditions fulfilling Bloch-wave conditions with complex wave number k in the direction of the waveguide and simple periodic conditions in the transverse direction~\cite{Davanco:2007}. As in the Bragg stack example we model gain by adding a small imaginary part $\epsilon''$ to the base material of the photonic crystal. For a specified real-valued frequency $\omega$ we find the associated complex $k$ by diagonalizing a complex matrix eigenvalue problem. Mathematically, changing the sign of $\epsilon''$ leads to the adjoint eigenvalue problem and thus the new eigenvalues are just the complex conjugates of the former. Physically, the group index and the PDOS thus remain unchanged when going from loss to a corresponding gain, while there of course is a change from a net loss to a net gain when inspecting the changes in $k''$.

To make contact to practical nanophotonic applications, we parameterize the homogeneous material gain as $g_0=2(\omega/c)n''$, where $n=n'+in''=\sqrt{\epsilon}$ is the complex refractive index of the material.
For the specific simulations we consider a semiconductor planar PhC ($\epsilon'=12.1$) with a triangular lattice of air holes, with lattice constant $a$ and air-hole diameter $d=0.5\times a$. Light is localized to and guided along a so-called W1 defect waveguide formed by the removal of one row of air holes from the otherwise perfectly periodic structure. Gain in such structures can be realized by embedding layers of quantum wells or quantum dots, which are pumped externally to provide net gain. For simplicity we restrict ourselves to a two-dimensional representation; this does not alter our overall conclusions. This PhC is known to support a guided mode, displaying a low group velocity when $k'$ approaches the Brillouin zone edge. In Fig.~\ref{fig:ng} we show the associated group index versus frequency. For the passive structure a clear divergence occurs around $\omega^* a/(2\pi c)=0.20525$. As $n''$ is increased the divergence is smeared out and eventually the group index approaches a constant value well below 50 throughout the frequency range for $n''$  still as small as  $7.2\times 10^{-3}$. Quite surprisingly, increasing the $n''$ from $1.4\times 10^{-5}$ by roughly a factor 500 to $7.2\times 10^{-3}$ causes a \emph{reduction} in the maximal group index from more than 500 to around 50. This shows that addition of gain may reduce the anticipated group index, and as a consequence, also the desired slow-light enhancement of the gain.

\begin{figure}[t!]
\includegraphics[width=0.5\textwidth]{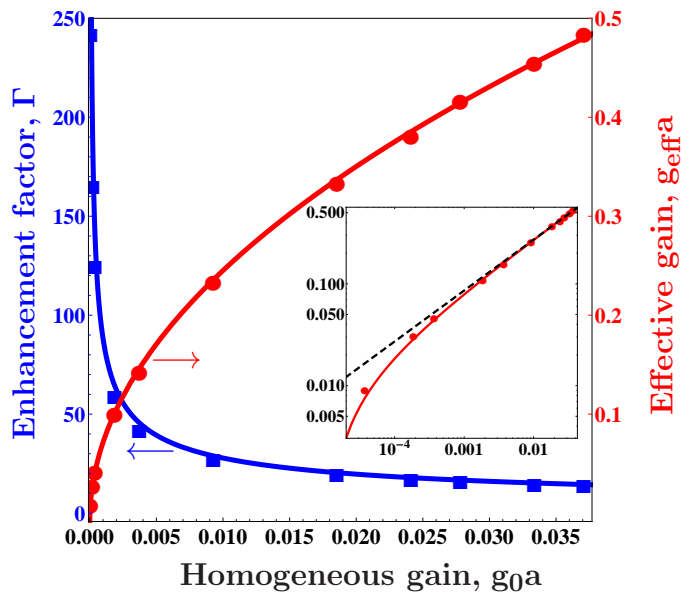}
\caption{(Color online) Slow-light enhanced gain $g_\textrm{eff}$ (right-hand axis) versus homogeneous gain $g_0$, evaluated at $\omega^*$ where the group index is initially maximal, see Fig.~\ref{fig:ng}. The red solid line shows a fit to the anticipated square-root dependence, Eq.~(\ref{eq:sqrtroot}), while the inset (log-log scale) exhibits minor quantitative deviations from a strict square-root dependence (dashed black line) due to a slight detuning from the band-edge singularity, see discussion in text.}
\label{fig:geff}
\end{figure}

Figure ~\ref{fig:geff} shows the effective gain $g_{\rm eff}=2k''$ (right-hand axis) versus $g_0$ evaluated at $\omega^*$ (where the propagation is initially slowest). Recalling the introductory discussion we anticipate an enhancement proportional to $n_g$ for low gain and indeed $g_\mathrm{eff}a$ starts out with a big slope in the low-gain limit, i.e. gain is greatly enhanced. However, at the singularity  $n_g(g_0)\propto g_0^{-1/2}$~\cite{Pedersen:2008}, and consequently
\begin{equation}
g_\mathrm{eff}(g_0) \propto n_g(g_0) g_0 \propto g_0^{1/2}\label{eq:sqrtroot}
\end{equation}
which is indeed supported by the full numerical data (red points) and the indicated square-root dependence (red line). The slow-light enhancement factor $\Gamma=g_\mathrm{eff}/g_0$ (blue line, left-hand axis) is correspondingly large for low $g_0$. Since $\omega^*$ is slightly detuned from the singularity a more detailed analysis yields $n_g \propto (\textrm{const.}+g_0)^{-1/2}$~\cite{Grgic:2010a} and consequently  a deviation from the square root dependence for small $g_0$  takes place (see inset). To make a connection with real gain materials, we consider an implementation at telecom frequencies with quantum dots as the active medium. Typically, $g_0$ is in the range of 10 -- $45\,{\rm cm}^{-1}$~\cite{Berg:2004} corresponding to $n''$ in the range from $1.5\times 10^{-4}$ to $7.5\times 10^{-4}$. The slow-light enhanced gain could then be as high as 1300 -- $2835\,{\rm cm}^{-1}$, corresponding to a gain enhancement extending from $\Gamma = 130$ down to 60 for the highest gain. This analysis implicitly assumes that the passive structure itself is ideal and with a diverging group index. However, disorder and imperfections will inevitably be present no matter the effort invested in the fabrication of the PhC. Ensemble averaging over disorder configurations will have the same overall effect on the PDOS as gain or absorption will have; singularities become smeared and the group index assumes a finite value. Clearly, such broadening can not be compensated by the addition of gain and the achievable effective gain may turn out lower than the estimate given above.

\emph{Symmetry points and Brillouin zone edges.} Finally, we discuss our results in the context of Bloch wave physics, inherent to the general class of periodic photonic metamaterials. From the Bloch condition, the dispersion relation $\omega(k')$ must necessarily be symmetric with respect to the zone edges (e.g., $k'=\pi/a$ for a Bragg stack). In the case of structures with zero gain (loss), this condition is met by $\partial\omega/\partial k'=0$ at the zone edge, corresponding to a standing-wave pattern.  However, in the presence of non-zero gain (loss), $k$ is in general complex and the mode may even propagate inside the bandgap region, albeit heavily damped. In this case, the symmetry condition is met by having two branches of solutions that extend across the bandgap and with a degeneracy at the zone edge (i.e. crossing bands near the center of the bandgap) and correspondingly the group index remains finite. Examples of such modes have been depicted in a number of recent works on lossy dielectric problems~\cite{Grgic:2011,Sukhorukov:2011} and  for damped plasmonic systems~\cite{Davanco:2007,Davoyan:2010}. In an attempt to compensate the inherent loss of metamaterials, gain should thus be added with care; while modes seem unaffected under a lasing condition (zero net gain) the anticipated dispersion properties may be jeopardized in an amplifier-setup if a too high net gain develops. We have focused on the regime of weak input signals, as appropriate to characterize the small-signal gain properties of an amplifier with no need to include saturation effects of the medium. Beyond this regime there would be a need for a self-consistent solution of the nonlinear light-matter coupling~\cite{Wuestner:2010,Hamm:2011}, possibly revealing new interesting findings when approaching the saturation regime.

In conclusion, adding gain to a periodically structured photonic material changes the dispersion properties and the slow-light enhanced gain in a complex manner. By both analytical examples and a numerical study we have illustrated how a large material gain degrades the slow-light properties supported by the corresponding passive structure, thereby eventually limiting the effective gain enhancement. Waveguide designs away from the band edge constitute an interesting case in the context of quantum-dot gain material. Here, the impact of gain is less detrimental and slow-light gain enhancement is possible with typical enhancement factors in the range from 60 to 130.

\emph{Acknowledgments.} This work was financially supported by the Villum Kann Rasmussen Foundation (via the NATEC Center of Excellence), the EU FP7 project GOSPEL, and the FiDiPro program of Academy of Finland.


\end{document}